\documentclass[conference]{IEEEtran}

\usepackage[table,xcdraw]{xcolor}
\usepackage{graphicx}
\graphicspath{ {img/} }
\usepackage[autosize]{dot2texi}
\usepackage[caption=false]{subfig}
\usepackage{tikz}
\usepackage{color}
\usepackage{multirow}
\usepackage{xspace}
\usepackage{pgfplots}
\usepackage{adjustbox}
\usepackage[utf8]{inputenc}
\usepackage{dirtytalk}

\usepackage[multiple]{footmisc}
\usepackage{fancyhdr}
\usepackage{flushend}

\newtheorem{definition}{\textbf{Definition}}

\newcommand{\ie}[0]{{\em i.e.},\xspace}

\newcommand{\eg}[0]{{\em e.g.},\xspace}
\newcommand{\etal}[0]{{\em et al.}\xspace}

\begin{document}

\title{Next Stop “NoOps": Enabling Cross-System Diagnostics Through Graph-based Composition of Logs and Metrics}

\author{\IEEEauthorblockN{Micha{\l} Zasadzi{\'n}ski\IEEEauthorrefmark{1}, Marc Sol{\'e}\IEEEauthorrefmark{1}, Alvaro Brandon\IEEEauthorrefmark{2},
Victor Munt{\'e}s-Mulero\IEEEauthorrefmark{1} and
David Carrera\IEEEauthorrefmark{3}}
\IEEEauthorblockA{\IEEEauthorrefmark{1}CA Technologies, Barcelona, Spain, \{michal.zasadzinski, victor.muntes, marc.solesimo\}@ca.com\\
\IEEEauthorrefmark{2}Universitat Politecnica de Madrid, Madrid, Spain, abrandon@fi.upm.es\\
\IEEEauthorrefmark{3}Universitat Politecnica de Catalunya, Barcelona, Spain, dcarrera@ac.upc.edu}}

\maketitle

\begin{abstract}
Performing diagnostics in IT systems is an increasingly complicated task, and it is not doable in satisfactory time by even the most skillful operators. Systems and their architecture change very rapidly in response to business and user demand. Many organizations see value in the maintenance and management model of NoOps that stands for No Operations. One of the implementations of this model is a system that is maintained automatically without any human intervention. The path to NoOps involves not only precise and fast diagnostics but also reusing as much knowledge as possible after the system is reconfigured or changed. The biggest challenge is to leverage knowledge on one IT system and reuse this knowledge for diagnostics of another, different system. We propose a framework of weighted graphs which can transfer knowledge, and perform high-quality diagnostics of IT systems. We encode all possible data in a graph representation of a system state and automatically calculate weights of these graphs. Then, thanks to the evaluation of similarity between graphs, we transfer knowledge about failures from one system to another and use it for diagnostics. We successfully evaluate the proposed approach on Spark, Hadoop, Kafka and Cassandra systems.
\end{abstract}


\section{Introduction}
Today's IT systems are large, dynamic, complex, and heterogeneous. The current and the future systems will frequently change their architecture and resources according to the business and user demand. Diagnosing them efficiently in satisfactory time (less than minutes) is already not within reach of even the most experienced operators. Because of that, the majority of trends and efforts around the development of troubleshooting and diagnostics of IT systems is driven by NoOps\footnote{http://cloudcomputing.sys-con.com/node/4054335/}\footnote{https://www.ibm.com/blogs/bluemix/2016/06/moving-devops-noops-microservice-architecture-bluemix/}\footnote{http://www.bmc.com/blogs/itops-devops-and-noops-oh-my/} business model~\cite{underwood2014death_ops}. NoOps stands for No Operations. One of the ways is the software automation. Then, it means a scenario of fully automated and self-manageable IT infrastructure. The shift of conventional operations to NoOps model is achieved by the full automation of maintenance activities, including failure diagnostics. In this model of maintenance, problems occurring in an IT system are solved immediately without any human intervention. 

However, to operate successfully in such a business model, the future diagnostic systems should perform precise, automated and fast root cause analysis. Also, these solutions should be able to diagnose problems even in a scenario where there is none or few data about failures and their causes. In many cases, recollecting the data necessary for diagnostics is expensive or even impossible. The use of similar data coming from another system with a different structure is a solution, but it is a considerable challenge. The solutions based on transfer learning can transfer and reuse as much knowledge on the behavior of a system as possible to keep pace with the changing architecture, infrastructure and rapidly growing number of knowledge domains.

So far, we have seen enormous work on automated diagnostics of IT systems, with use of data mining or Artificial Intelligence (AI)~\cite{kwon2017survey_deep,anomaly_detection_2016_svm_dl}. Most of this work uses for diagnostics either metrics or logs. When both are used, the use of logs is limited to counting specific key terms or entries with a specific severity level. Another common limitation of current systems is the lack of inclusion of detailed system information, \ie connectivity, hardware specification in diagnostics. There is still room for improvement in knowledge integration and knowledge transfer before we reach the era of NoOps. As we show in this publication, integrating log entries, metrics, and other system data improve the accuracy of the diagnostics for IT systems. 

In this paper, we propose a cross-system root cause classification framework based on similarity evaluation of weighted graphs with multi-attribute nodes. The framework uses logs, metrics, configuration and connectivity information to represent the state of a system as a graph. Then, the framework evaluates the similarity between an abnormal state and a collection of previously diagnosed states. By finding the most similar graph in the solution space, we can classify the anomaly and provide a root cause. Moreover, we use automatically calculated weights to highlight the system metrics that better describe a failure. Finally, we use the framework for a cross-system failure classification. By acquiring a collection of diagnosed anomalies for one system architecture, we can establish the root cause for anomalies that occur in a completely different architecture (cross-system diagnostics). 

Rapidly changing system architecture is a consequence of new requirements and scaling of a system. We leverage the proposed framework in this scenario. Using knowledge transfer, just after starting a new architecture of a system, we can diagnose it and proactively avoid failures. The proposed system does not only allow for precise diagnostics but also helps in proactive avoidance of failures. The system can output the nearest possible future failure as a result of graph similarity evaluation. Such an approach, saves time, effort and results in performance and reliability advantage over competitors.

We evaluate the proposed framework in the environments running representative and different Big Data applications such as Spark~\cite{spark} and Hadoop~\cite{hadoop}. We inject failures into these environments and evaluate the quality of failures classification, reaching more than 70\% of both \textit{f1-score} and accuracy. Then, we perform experiments using different architectures with containers running Cassandra~\cite{cassandra} and Kafka~\cite{kafka} systems. We evaluate our cross-system nearest root cause classification when the symptoms of failures are known only for one of these systems. We receive average \textit{f1-score} 77\% with the same level of accuracy.

The remainder of this paper is divided into seven sections. Work related to graph-based root cause analysis systems, cross-system knowledge transfer and use of logs for diagnostics is discussed in Section~\ref{section_related_work}. In Section~\ref{section_background} we describe the background for the graph similarity calculation. Then we present the framework for creation and similarity evaluation of automatically weighted graphs representing a system's state that contains: metrics, logs, system connectivity, infrastructure. Our contributions are:
\begin{itemize}
 \item A solution on how to include logs in a graph representation of a system state. (Subsection~\ref{subsection:logs})
 \item A method for automatic adjustment of weights of nodes and node attributes, according to the distribution of a metric. (Subsection~\ref{subsection:automatic_importance})
 \item Evaluation of the proposed solution on real datasets for root cause classification. This Section presents an evaluation of the proposed framework on a cluster running Hadoop and Spark jobs. We prove that including logs and the automatic importance assignment system increases the accuracy of the classification with respect to other methods. (Section~\ref{section_evaluation_method})
 \item Evaluation of root cause classification in cross-system transfer learning; We search for a failure using knowledge captured from one system (Kafka) and utilize it in another system (Cassandra). We prove that the graph approach can transfer knowledge to/from Cassandra from/to Kafka. (Section~\ref{section_evaluation_cross_domains})
\end{itemize}

Both evaluation sections contain results from four use cases running in different infrastructures: on-premise cluster and containers in a cloud. This strategy allows us to prove the reproducibility and broad usability of the proposed framework. We conclude the paper with the discussion and plans for future research in Section~\ref{section_discussion}.

\section{Related work}\label{section_related_work}

\subsection{Graph-based systems for root cause classification}
Monitoring and logging systems are responsible for providing full observability of a system state, which is one of the most important inputs for a root cause classification system. Current research in this field is focused on dealing not only with the huge size and complexity of information encoded in logs but also with fault tolerance and use of partial information~\cite{pivot_tracing2015}. Usually, operators use these two sources of information in a troubleshooting process separately, and diagnostic tools do not combine well descriptive data with metrics. One of the best ways to do so is utilizing a graph representation of a system state. Constructing proper graph representation allows for anomaly detection and diagnostics~\cite{graph_anomaly_detection_2015}. Graph-based approaches are widely used for root cause classification, detection and prediction of abnormal events and failures~\cite{bayesian_failure_prediction_2015,sdn_failure_detection_2016,anomaly_detection_graph_series_2017,noble2003graph,log_data_graph_db_2018}. 

\subsection{Cross-system failures knowledge transfer through similarity evaluation}
Diagnostic systems can gather the knowledge in one domain and reuse this knowledge for diagnosing similar systems with symptoms in similar knowledge domains. Generally, we call this type of use of knowledge \textit{transfer learning}~\cite{transfer_learning_2015} and the \textit{heterogeneous transfer learning} when the knowledge comes from different systems~\cite{pan2010survey_transfer_learning}. In this paper, we focus on a scenario of the \textit{transductive transfer learning}, where the data is labeled in the source domain, but not in the target knowledge domain. According to this area, one of the paths to deploy transfer learning in diagnostic systems is to apply similarity measures between a diagnosed state and the abnormal state to be diagnosed. When we represent system states as graphs, we can compare the transferred state by evaluating similarities between them~\cite{zager2008graph_sim,koutra2011algorithms}. The work of Papadimitriou \etal~\cite{papadimitriou2010web_graph_sim} is an important contribution in the field of diagnostics via graph similarity. This work evaluates graph similarities to find anomalies in the web. The authors consider different approaches for similarity evaluation which are limited to the topologies of compared graphs. In comparison, in our approach for the cross-system diagnostics, we encode more information. We use attributes of different types in both edges and nodes, together with the information contained in system and application logs, providing a much more detailed input for the graph similarity function. Work on the similarity between different texts and logs is presented in~\cite{gomaa2013survey_text_sim,islam2008semantic} and it is widely used for diagnostics of IT systems. Research of Putra \etal~\cite{putra2017evaluating_text_sim} includes graph-based text similarity evaluation. Other important work on utilizing similarity between logs that are used for diagnostics can be found in~\cite{li2017mining,Flesca2005,microsoft2016}.

\subsection{Mining logs for root cause classification and diagnostics}

The logs of an IT system are a valuable source of information used for data-driven diagnostics and prognostics of a system state. A usual method of working with logs, it is exploring the statistics and the occurrence of a set of key terms using log parsers, indexers, and miners. The authors of a survey on data-driven techniques in computing system management~\cite{Li:2017:DTC:3101309.3092697} claim that to realize the goal of self-management, systems need to automatically monitor, characterize, and understand their behaviors and dynamics; mine events to uncover useful patterns, and acquire valuable knowledge from historical log/event data. Fundamental knowledge of diverse approaches of error log processing is found in~\cite{salfner2008error,log_preprocessing_2009}. Some simple mining methods include log key terms occurrence correlation~\cite{logan2016}, and modeling a multithreaded system behavior through graphs or sequences representing system calls. For instance, the authors of~\cite{salfner2008error} deal with the problem of failure prediction through clustering similar log sequences. They propose an algorithm to assign the source of failures to logs, using Levenshtein’s edit distance. 

Recently, a considerable part of work on automated diagnostics is performed with the help of AI. DeepLog~\cite{deeplog_2017} is one of the most significant contributions. The authors propose a system for anomaly detection and diagnosis which is based on deep learning. The performance and accuracy of the solution are high. However, to use it, there is a necessity of defining metadata. For this reason, the solution has limited usability regarding full automation. Authors of~\cite{logsed_ibm_2017} propose an approach to mine time-weighted graphs from logs with many threads running. The solution evaluated on the cloud environment performs with high \textit{f1-score} that is about 80\%. Authors in~\cite{logs_causal_mining_2017} use casual inference to diagnose network failures. They mine casual graphs from logs, considering connected devices in a graph. One of the conventional approaches to deal with log preprocessing and comparison is transforming log entries to vectors, using the Word2Vec algorithm~\cite{word2vec_Mikolov:2013,word2vec_explained14}. 
A recent attempt to leverage Word2Vec for root cause classification is described in~\cite{logs_nlp2017}. The authors propose a method for processing logs with a Word2Vec model and then using a Bayesian classifier.

Analyzed state of the art shows the dependency between accuracy and the underlying complexity of the solutions. Much state of the art research is focused on accurate analysis and mining of logs based on metadata for a specific log structure. There are not many solutions which diagnose a system just by consuming logs without specific preprocessing techniques. With the solution that we propose in this paper, we would like to fill this gap. The solution is as general as possible, and it can work with many IT system types with little human effort to deploy the framework that uses logs, metrics and others system information data.

\section{Background: Graph Similarities}\label{section_background}
In this section, we provide background knowledge on the problem of similarity calculation between graphs. We define the problem, the graph representation, and how to calculate similarities between different node attribute types.

\textbf{Similarity.}  
According to~\cite{koutra2011algorithms}, we define the problem of finding a similarity between two graphs as follows.
\begin{definition}
Given two graphs $G_1(V_1, E_1)$, $G_2(V_2, E_2)$. Find an algorithm to calculate the similarity $s$ of the graphs, which returns a number between 0 and 1. Two graphs have similarity $s=1$ only when they are identical while a similarity value of $0$ intuitively says that they are completely different. \end{definition}

\subsection{Approximate Graph Similarity Calculation}\label{subsection:approx_graph_repr_calc}

\textbf{Graph representation of a system state.} Graphs allow representing an IT system state including all types of data which can describe that state. A graph is defined as a set of $\{E, V, W, A, S\}$ corresponding to the sets of \textit{edges, vertices, weights, attributes, and similarity functions}. Each of the system components is a node that has multiple attributes and represents a different level of abstraction, \eg hardware, server, an application, application module, application thread, container, or a microservice. Edges represent the connectivity between system components. Attributes of a node contain different information encoding the system state, \eg metric value, log entries, component type, software details.

Also, to represent the different importance of each of the attributes, we introduce \textbf{weights} at each level of the graph structure. We use them with each element of a graph: edges, nodes and node attributes. A weight indicates how significant is the influence of the similarity between particular elements on the final similarity result. Primarily, an expert can define weights through the root cause analysis framework. When an anomaly is detected inside the system state graph, the expert can pinpoint the metrics and components that are more important inside that anomaly. These will be later used as inputs by the root cause classification system. Such an intuitive mechanism creates permanent opportunities for the framework to gather the expert knowledge. In Section~\ref{section_contribution}, we introduce the automatic weight calculation mechanism to deal with limitations of the manual weight assignment.

\textbf{Graph similarity calculation.} We calculate the maximum similarity $s(G_1,G_2)$ between two graphs $G_1=(E,V,W,A,S)$ and $G_2=(E,V,W,A,S)$. It is a well-defined optimization problem, which consists of matching a node of a graph with the most similar one of the other graph. We use \textit{hill climbing}~\cite{Sorlin:2005:RTS:2140606.2140626} to solve the optimization problem. The similarity between two nodes is calculated by using their attributes, which can be both logs and metrics. In order to do that, we specify importance of each attribute - a \textbf{weight}, and a \textbf{similarity function}. We use weights for calculation of the weighted average similarity between attributes. In the Subsection~\ref{subsection:attributes} we propose different similarity functions depending on the attribute types - custom functions to compare different elements of a graph.

\subsection{Similarity between different attribute types}\label{subsection:attributes}
We define similarity functions for numerical, vector, categorical and ontological attributes in Table~\ref{table:sim_functions}. Because of using different similarity functions, we manage the calculation of similarities between different attribute types, coming from the two compared graphs.
\begin{table}[ht]
\centering
\caption{Similarity functions used in the proposed framework}
\label{table:sim_functions}
\resizebox{0.48\textwidth}{!}{%
\begin{tabular}{|l|l|}
\hline
\textbf{Type of attributes} & \textbf{Similarity function} \\ \hline
Numerical & \begin{tabular}[c]{@{}l@{}} $1-scaled\_distance(a_1,a_2)$\end{tabular} \\ \hline
Vector & $cos(a_1,a_2)$; inverse Euclidean distance; Minkowski $p$ distance\\ \hline
Categorical &  $1\ if\ a_1==a_2\ else\ 0$ \\ \hline
Ontological & \begin{tabular}[c]{@{}l@{}} Modified Wu and Palmer \cite{wu1994} similarity metric \\ $\frac{2\cdot d(C)}{d(C1)+d(C2)}$ \end{tabular} \\ \hline
\end{tabular}%
}
\end{table}

\noindent \textit{Similarity between numerical attributes}. This function is used for those metrics that take numerical values such as CPU usage, bytes written to disk or memory used to name a few. More specifically, for numerical type attributes $a_1$ and $a_2$, we use the formula $s(a_1,a_2)=1-\frac{|a_1-a_2|}{|max-min|}$. Two points that are close on the scale, will have a higher similarity value. They achieve the maximum similarity $=1$ only if they are equal.

\noindent \textit{Similarity between vectors.} Vectors can represent a measurable state of a system module, but can also represent text inside a log file, as we will explain in Subsection~\ref{subsection:logs}. The similarity between vectors is usually defined by the value of cosine between two vectors. Also, other metrics that are based on different distance formulas can be used.

\begin{figure}[ht]
 \includegraphics[width=0.2\textwidth]{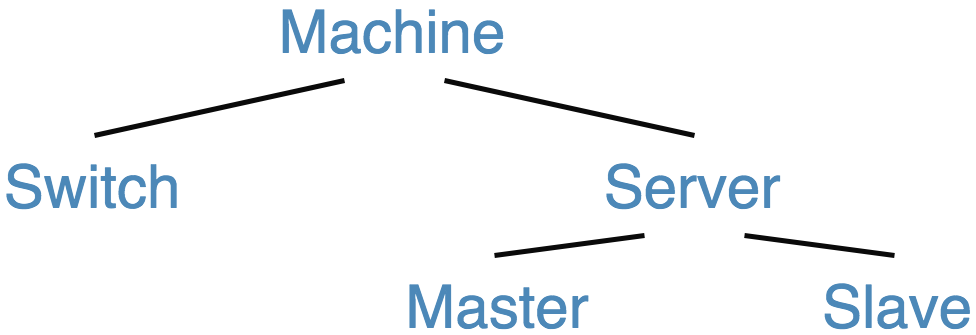}
 \centering
 \caption{An example taxonomy defining equipment type used in the evaluation. For instance, using the ontological similarity formula from Table~\ref{table:sim_functions}: $similarity(Master,Slave)=0,66$, $similarity(Master,Switch)=0,4$, $similarity(Server,Switch)=0,5$ }
 \label{fig:taxonomy}
\end{figure}

\noindent \textit{Similarity between types.} Graph nodes contain attributes which specify a type. A taxonomy is a tree that represents a hierarchy of concepts in a given domain. In Figure~\ref{fig:taxonomy}, we present an example taxonomy. Each node in a graph can contain attributes that define its type inside this taxonomy. The functions used for similarity calculation between different types are introduced in Table~\ref{table:sim_functions}.

\noindent \textit{Similarity between categories.} They take values that are names or labels, \eg the image of a Docker container (\eg Haproxy, WordPress), disk label, hardware model. According to the categorical values, the similarity is $1$ when the values are equal, otherwise $0$.

\section{Weighted graphs representing system state for cross-domain diagnostics}\label{section_contribution}
\begin{figure}[ht]
 \includegraphics[width=0.48\textwidth]{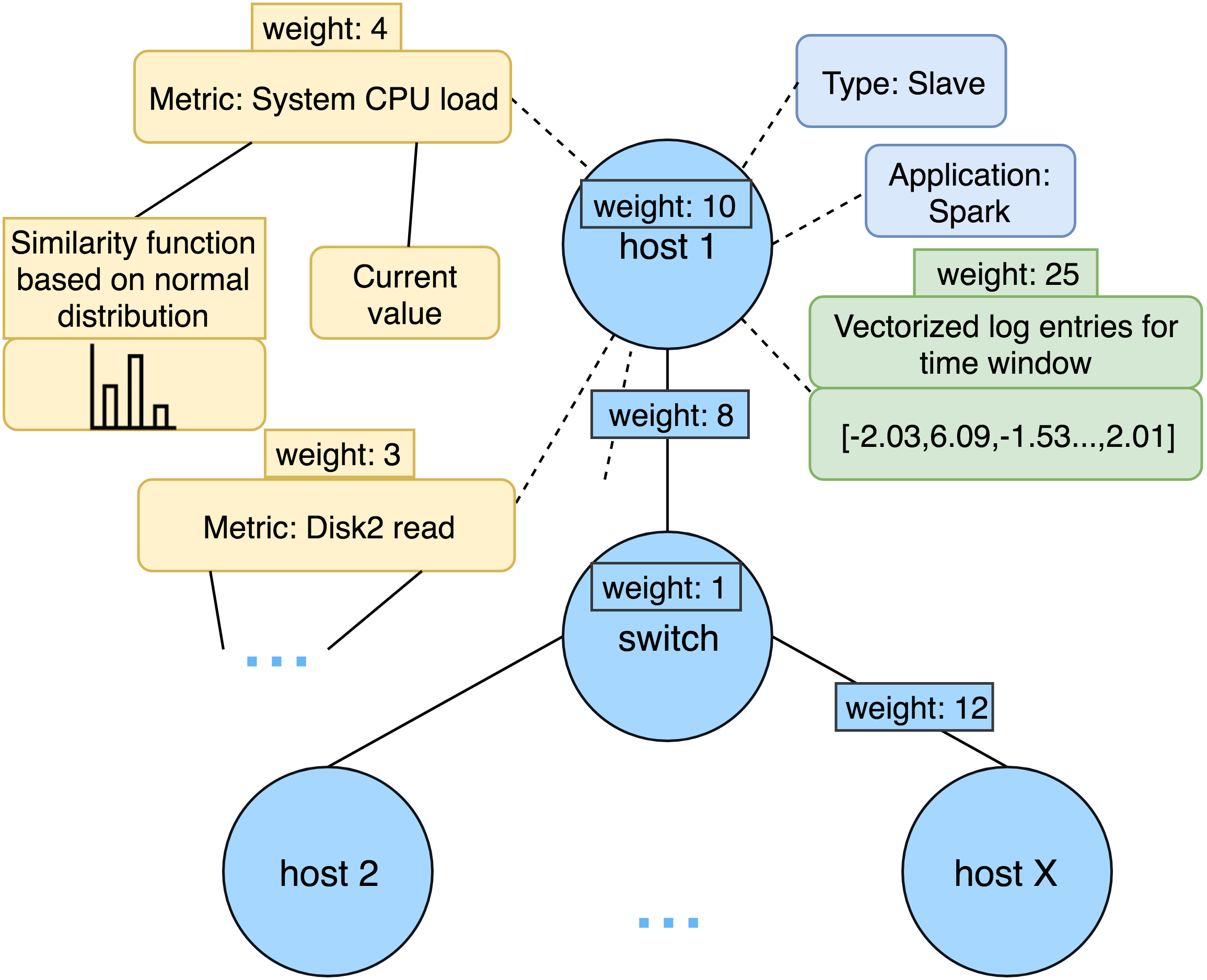}
 \centering
 \caption{An example graph with multi-attribute nodes representing a system state, including connectivity between devices their types, metrics, and logs. Each node contains many attributes, which are different types: categorical, numerical, vector, distribution, classification.}
 \label{fig:example_graph_representing_system}
\end{figure}

Motivated by the challenge of shifting operations to NoOps, we present the following contribution. First of all, we propose a diagnostic framework based on an automatic similarity calculation for graphs representing a system state. The framework automatically adjust graph weights according to the distribution of historical values of metrics. Also, the weight module allows for adjusting the importance of a metric according to an operator's feedback. Weights are used to indicate the important elements of a system which hold significant information for diagnostics. For instance, in case of a network failure, attributes with network-related metrics will be more important than those non-related, \eg CPU, temperature. The framework reacts to a trigger based on anomaly detection mechanisms, \eg an error message, exceeded the threshold of a metric. It outputs the similarity score between the current state of the system and previously acquired anomalous states. Such information can be used for early detection of failures and their prevention. In Figure~\ref{fig:example_graph_representing_system}, we present an automatically weighted graph representing a system state. Blue nodes represent system elements, in this case: hosts, and a switch. Each node contains many attributes which can contain static information, \eg node type, and runtime data, \eg metric values, metric distributions.

\begin{figure}[ht]
 \includegraphics[width=0.48\textwidth]{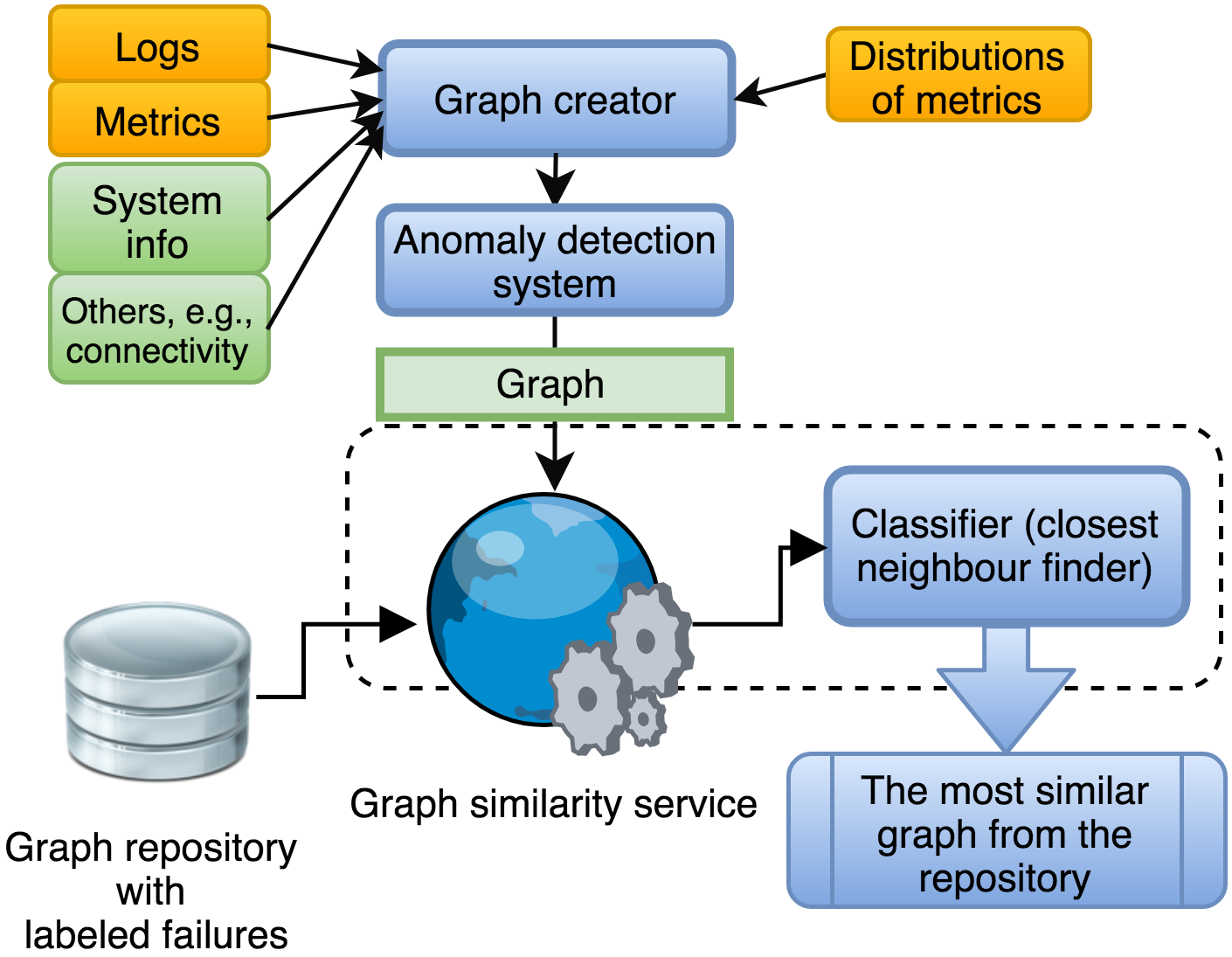}
 \centering
 \caption{Scheme presenting the architecture of the root cause classification framework working with an external anomaly detection system.}
 \label{fig:system_architecture}
\end{figure}

In Figure~\ref{fig:system_architecture}, we present the proposed framework for root cause classification. The framework manages the creation of weighted graphs and calculation of similarity between them. One graph comes from a repository with anomalous graphs that have been previously labeled with its root cause, and the other one represents an anomalous state of a diagnosed system. Note that we assume the existence of an anomaly detection system that can extract anomalous system graphs. 
Usually graphs are labeled automatically by an anomaly detector. Also, when it is necessary an expert can label them. The graph creator builds graphs that represent the system state. They use sources of data coming from different monitoring systems or other information about the system architecture. The content of graphs and their topology depends on the modeling approach. For instance, each node can represent a server, application or its module. The graph similarity module is used to find in a solution space the nearest graph to the anomalous system state graph. By finding this closest labeled graph, we can know the cause of a failure. In case of use of the proposed framework for failure prevention, we get a graph representing the most probable failure which is likely to occur.

\subsection{Including log data}\label{subsection:logs}

In this subsection, we propose a log representation structure that can be embedded into our graph. In the proposed graph representation of a system state, the attributes capture information from different sources, including logs. In contrast to many state of the art solutions, we consume logs without any metadata or dependency on its structure. Thanks to this approach, our solution is agile and needs a minimal effort for the deployment. We only extract timestamp, severity level, and the rest is treated as a log entry that includes application module name, message, thread name, and other fields. 
Moreover, users (framework operators) can disassemble logs by modules and put them inside new nodes or attributes representing these modules in a system state graph. For instance, an operator deploying the proposed framework may decide that graph representation of a system should be a detailed one. Then, a node presenting a host is connected with its child nodes, representing some modules, \eg threads, classes, application modules. Logs of this host are split among these nodes.

\begin{figure}[ht]
 \includegraphics[width=0.3\textwidth]{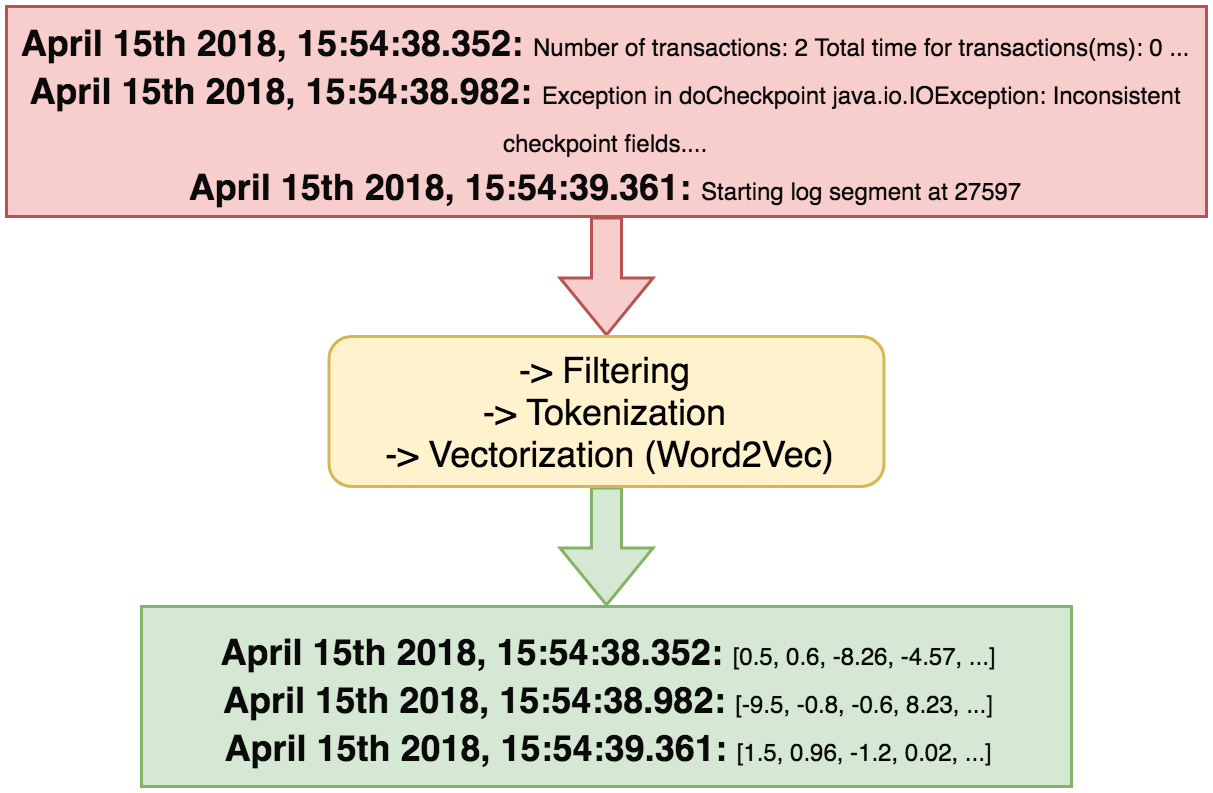}
 \centering
 \caption{An example process of transformation log entries to vectors.}
 \label{fig:word2vec_example_processing}
\end{figure}

We propose to use vectorized logs using Word2Vec models, in a system's state representation for the following reasons. The whole log processing is a simple algorithm and includes removal of special chars, sequences, and stop words, tokenization and vectorization. The scheme illustrating the whole process is presented in Figure~\ref{fig:word2vec_example_processing}.\par \textit{Filtering.} After eliminating special char sequences \eg hex strings, the vocabulary in logs is limited. Typically, human-created templates of logs do not contain synonyms, just strict and simple phrases. After this stage, log entries contain less noise and represent a state of the generalized system, rather than a particular case. Also, removing special characters helps to avoid model over-fitting. This step does not only improve the model quality but also transforms a log into a universal form, which is mandatory in cross-system diagnostics. \par
\textit{Tokenization}. The tokenization step disassembles sentences into bags of words. \par
\textit{Vectorization.} Thanks to Word2Vec we transform log into vectors. The vectorization stage enables to represent log entries in relatively small models, which we show later in the evaluation in Section~\ref{section_evaluation_method}. Firstly, it is necessary to create a model mapping the vocabulary into $n$ dimensional space. The performance of a model depends on its configuration parameters and the size of the vocabulary used for training. A considerable advantage of using a Word2Vec embedding model is that it performs well even if it is trained using the vocabulary of one domain and used for another. Also, similarity calculations should be as fast as possible to enable diagnostics of failures in a dynamic environment. Hence, it is not feasible to use natural language processing (NLP) techniques such as key terms extraction using rank algorithms for each log sentence as we demonstrate in Section~\ref{section_evaluation_method}, where we test different approaches. The proposed log processing algorithm does not need much configuration work. We only need to adjust a time window size, which starts with a specific severity type. In our case, we propose to use severities with a higher level than the warning one.

\begin{figure}[ht]
 \includegraphics[width=0.25\textwidth]{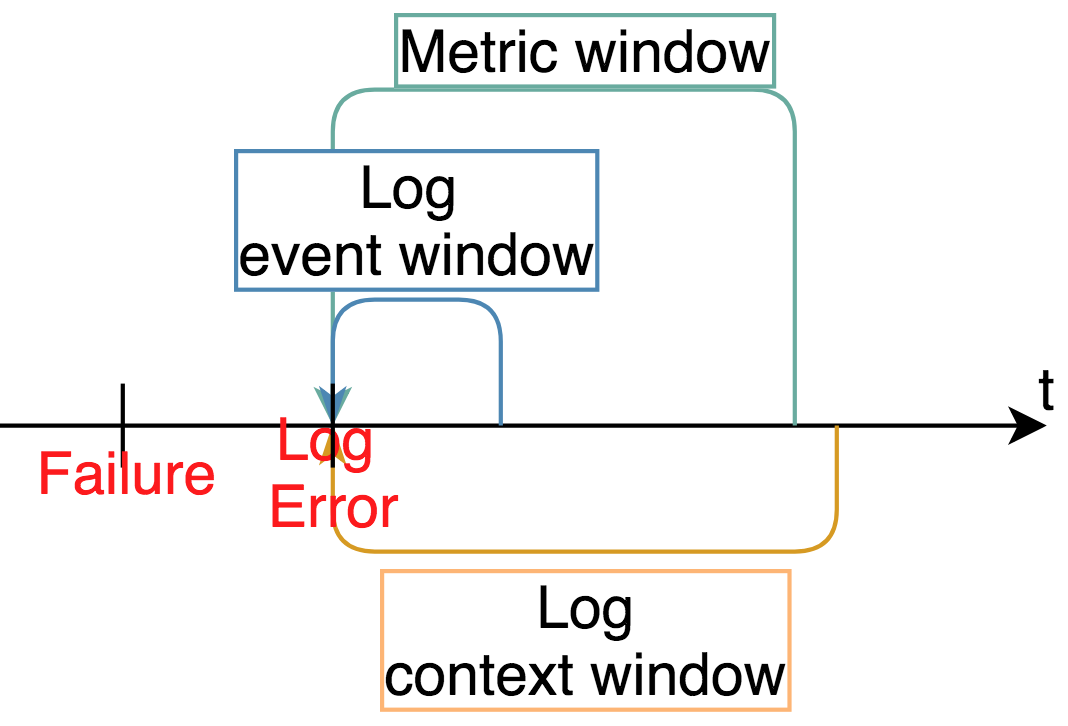}
 \centering
 \caption{Scheme presenting context and event window of logs. Both windows start on a first Error or Warning message.}
 \label{fig:time_explained}
\end{figure}

After a failure occurs, we can find messages on the logs containing information for that failure, while some others are just messages belonging to the usual operation of the system components. As discussed in~\cite{word2vec_windows_discussion_levy2014dependency}, using smaller time windows capture the more detailed meaning of a word (in our case, if it mentions a failure), and large ones which capture the context (general context of the application used). We propose to use two log windows: one called (1) \textbf{context window}, and the other one (2) \textbf{event meaning window}. The context window represents the general state of a part of the system. Mainly, it enables to capture application's normal activities. The event meaning window captures log entries in a shorter time after a particular event. Logs in such a window represent specific information about the event. Both windows start when an error or warning message is written into the log. The reason we take this approach is that operators usually do not know when the system starts failing, but they know the precise time of every error or warning written to logs. We explain the concept of window lengths in Figure~\ref{fig:time_explained}.

\subsection{Using metrics distribution for automatic weighting of node attributes and measuring similarity}\label{subsection:automatic_importance}

The distribution of values for a metric can be used to know how different or uncommon is a value observed in the system. In this subsection, we explain how we use the cumulative distribution function to our advantage by, firstly, calculating weights automatically inside the graph representation, and secondly, comparing two numerical attributes taking into account the distribution of their historical values. 

\subsubsection{Automatic weighting of node attributes}
There are two ways of defining weights in graphs which represent the importance of the different elements of the system status.

\textbf{Firstly}, thanks to the weight assignment mechanism, operators adjust the importance of a particular metric in the graph representation based on their expert knowledge of a failure. For instance, operators might put a higher weight on the CPU load than on the disk IO, for a problem related to a system overload. Thanks to this approach, we do not require from operators to know specific characteristics or deviations of metrics. We use a part of their expertise which contains importance of metrics used in a troubleshooting process. However, for many systems this type of assignment can be impractical, \eg complex system, weights get outdated.

\textbf{The second possibility} for weight assignment in graphs is an automatic weight calculation from available metric data. 

In this subsection, we focus on the latter. We propose an \textbf{automatic weight assignment} mechanism to automatically assign the importance of an attribute, given its distribution. 

According to the troubleshooting activities of IT operators, the more abnormal the attribute value is, the better describer of a particular failure. In this case, we define weights which are proportional to the deviation of the usual value for an attribute distribution of values. For instance, using the \textit{normal distribution} $X \sim \mathcal{N}(\mu,\,\sigma^{2})$ where $\mu$ stands for the metric mean value, and $\sigma$ stands for standard deviation, we have the following definition.
\begin{definition}
The weight of a numerical attribute which is proportional to the deviation of a metric value $a$ is defined as $w(a)= \frac{ |a-\mu| }{ \sigma }$
\end{definition}

\subsubsection{Measuring similarity from metric distributions}
The similarity function based on metric distribution enables to utilize data containing historical values for an attribute. The function definition contains cumulative distribution function (CDF) and its parameters. We define the similarity function between two numerical attributes, as the formula $similarity=1-distance$ where $distance$ is the difference between $CDF$ values of attributes. For the normal distribution used in the proposed framework, we have the following definition. 
\begin{definition}
Given numerical attributes $a_1, a_2$ from two graphs and distribution of these attributes $X \sim \mathcal{N}(\mu,\,\sigma^{2})$, where $\phi$ stands for the CDF of this distribution, their similarity is provided with the formula $similarity=1-|\phi_{\mu,\,\sigma^{2}}(a_1) -\phi_{\mu,\,\sigma^{2}}(a_2) |$
\end{definition}

The above two simple mechanisms allow to automatically include the importance of attributes in the graph representation of a system state and similarity calculation.

\subsection{Enabling cross-system diagnostics}\label{subsection:cross_domain}

\begin{figure}[ht]
 \includegraphics[width=0.48\textwidth]{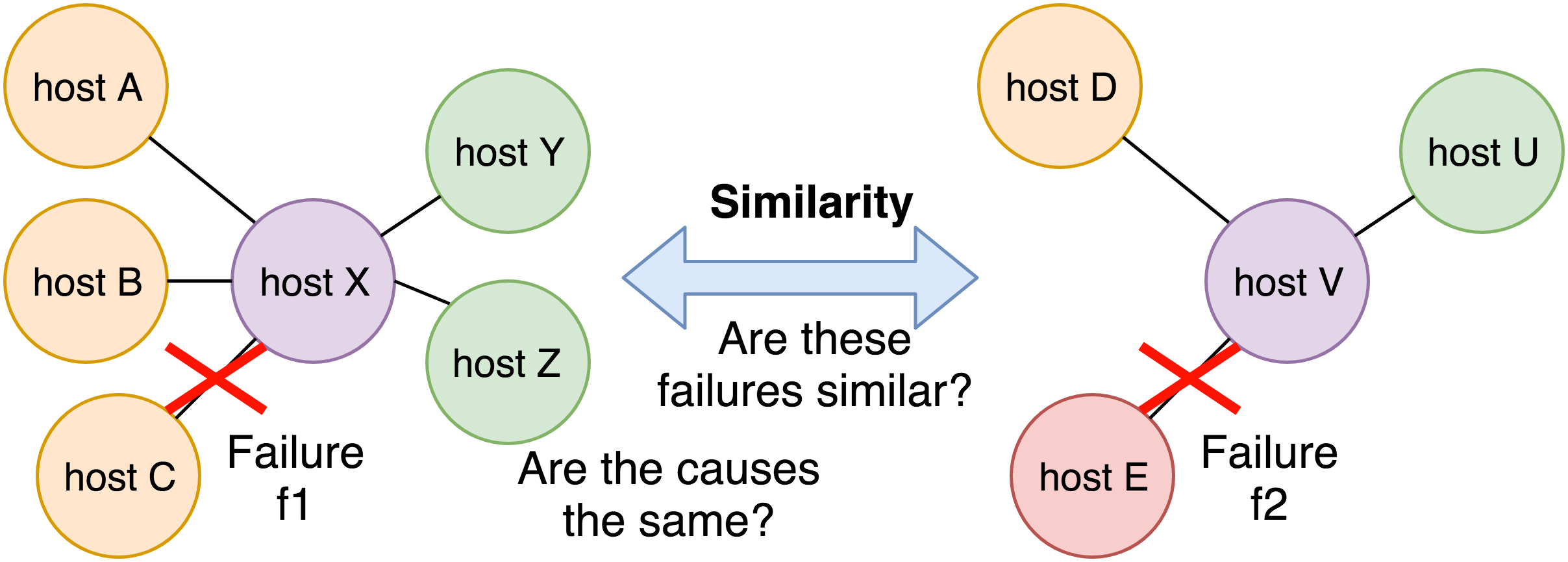}
 \centering
 \caption{Scheme illustrating an idea of cross-system graph comparison.}
 \label{fig:scheme_cross_domain_failuresl}
\end{figure}

Finally, we use the proposed framework to transfer knowledge about failures from one system that we call \textit{source system} to another that is called \textit{target system}. In Figure~\ref{fig:scheme_cross_domain_failuresl}, we present the cross-system knowledge transfer problem. A source system and a target system can have both different topologies and contents of nodes. We use the proposed graph representation of system states as a medium to transfer knowledge about failures. Then, thanks to the framework, we can compare two states of different systems and calculate the maximum similarity of these states. In the final step, we find the nearest graph, which best describes a target system state by knowledge coming from a source system. 

In details, using our framework, knowledge transfer is possible because of:
\begin{enumerate}
 \item Calculation of the maximum similarity between two graphs with different structures using different similarity functions (Subsection~\ref{subsection:attributes}). The framework finds the maximum similarity by matching proper subgraphs. Also, defining a taxonomy allows for the calculation of the similarity between two nodes that are different but represent the same concept in a domain. For instance, a slave server of Spark and a data node of Hadoop are close to each other inside the taxonomy, because they are both slaves in a master-slave architecture. 
 \item Inclusion of logs in the graph representation, as they describe in natural language events that happen in the system, independently of their architecture or resource usage (Subsection~\ref{subsection:logs}). The two log windows (context and event) contain universal descriptive information, no matter what the differences are between the topologies and components of the two system graphs.
 \item Including the information contained in the distribution of the metrics for a given architecture. We do it through the automatic weight assignment and the similarity function based on the distance between distributions (Subsection~\ref{subsection:automatic_importance}). The metric values registered for the source and target system can be very different depending on their resource usage patterns. Calculating weights and measuring the similarity using their distributions, allows for a better comparison between two different systems. 
\end{enumerate}

\section{Evaluation and experiments: Root cause classification}\label{section_evaluation_method}
In this Section, we show through a series of experiments the quality of our proposed root cause classification framework. We evaluate different features of the framework and compare them to representative and popular state of the art techniques. We use a \textit{f1-score} metric which both includes recall and precision. In this Section, we evaluate the quality of the framework in a scenario where the source and the target system is the same. For this task, we use two use cases: a Spark cluster and a Hadoop cluster. We evaluate cross-system diagnostics in Section~\ref{section_evaluation_cross_domains}, using Kafka and Cassandra systems.

\subsection{Experimental methodology}

\textbf{Experimental environment}. In the first set of experiments, we use the following experimental system to create the dataset. The system comprises:
\begin{itemize}
 \item 5x amd server: 32GB RAM, AMD Opteron(tm) Processor 6168 (12 cores, 1.9 GHz), equipped with IPMI card and running Ubuntu OS
 \item Switch D-link DGS-1210-48
 \item 2x Power Analyzer ZES Zimmer LMG450. The device is a 4 Channel power analyzer mounted in a rack and connected between each power supply and servers and the switch.

\end{itemize}

\newcommand*\justify{%
  \fontdimen2\font=0.4em
  \fontdimen3\font=0.2em
  \fontdimen4\font=0.1em
  \fontdimen7\font=0.1em
  \hyphenchar\font=`\-
}

\renewcommand{\texttt}[1]{%
  \begingroup
  \ttfamily
  \begingroup\lccode`~=`/\lowercase{\endgroup\def~}{/\discretionary{}{}{}}%
  \begingroup\lccode`~=`[\lowercase{\endgroup\def~}{[\discretionary{}{}{}}%
  \begingroup\lccode`~=`.\lowercase{\endgroup\def~}{.\discretionary{}{}{}}%
  \catcode`/=\active\catcode`[=\active\catcode`.=\active
  \justify\scantokens{#1\noexpand}%
  \endgroup
}

The monitoring system acquires 22 metrics representing the system state, such as CPU total load: \texttt{idle, iowait, softirq, system, user;} disk: \texttt{bytes read, bytes write, IO read, IO write;} memory: \texttt{buffer cache, free, map, used;} network: \texttt{received bytes}, \texttt{received packets, send bytes, send packets,} and processes: \texttt{load10, load15, load5, number of running processes}. The power meters acquire energy consumption of the servers and the switch. The probing period is set to 5 seconds. The monitoring system works on InfluxDB\footnote{https://www.influxdata.com/} stack and we use ElasticSearch\footnote{https://www.elastic.co/} stack for log storage.

\textbf{Workloads}.
During the experiments, we generate Hadoop and Spark workloads using HiBench~\cite{hibench}. We use workloads such as sort, word count, k-means clustering, Bayesian classifier. Each workload takes from 20min to 2h. Random workloads run continuously. 

\subsection{Failure and anomaly injection}
We inject different failures in the experimental environment. Each of the described failures is injected 20 times. We choose a set of failure types which are representative and well-aligned with use cases in real environments. Also, different failures should manifest exclusive symptoms in different metrics and logs. The next criterion of choosing the failure types is that they should differentiate possible scenarios of lacking data that are often caused by connectivity problems. 

The following list presents the injected anomalous workloads and failures.
\begin{itemize}
\item \textbf{High CPU load}. Background process running CPU pattern of 100\% load for 90\% of server cores. This failure simulates a scenario of a node slow-down, caused by \eg an unfinished job, unwanted or unfinished process. CPU performance degradation can also simulate a failure of one of many workers in a Big Data cluster.
\item \textbf{High disk load}. Random write and read operations on a 10GB file, generated with the FIO utility\footnote{https://github.com/axboe/fio/}. This failure simulates a scenario of a failed disk in a disk array. Thanks to this failure type we can observe many HDFS errors.
\item \textbf{High network transfer}. 20 threads are uploading and downloading 5GB files. It simulates significant network slowdowns, which can occur as a result of network infrastructure failure.

\item \textbf{Host shutdown}. Immediate node shutdown through IPMI card. It simulates a node crash, a sudden and unexpected failure of the whole machine.

\item \textbf{Network failure}. Physical disconnection.

\end{itemize}

\begin{figure}[ht]
 \includegraphics[width=0.4\textwidth]{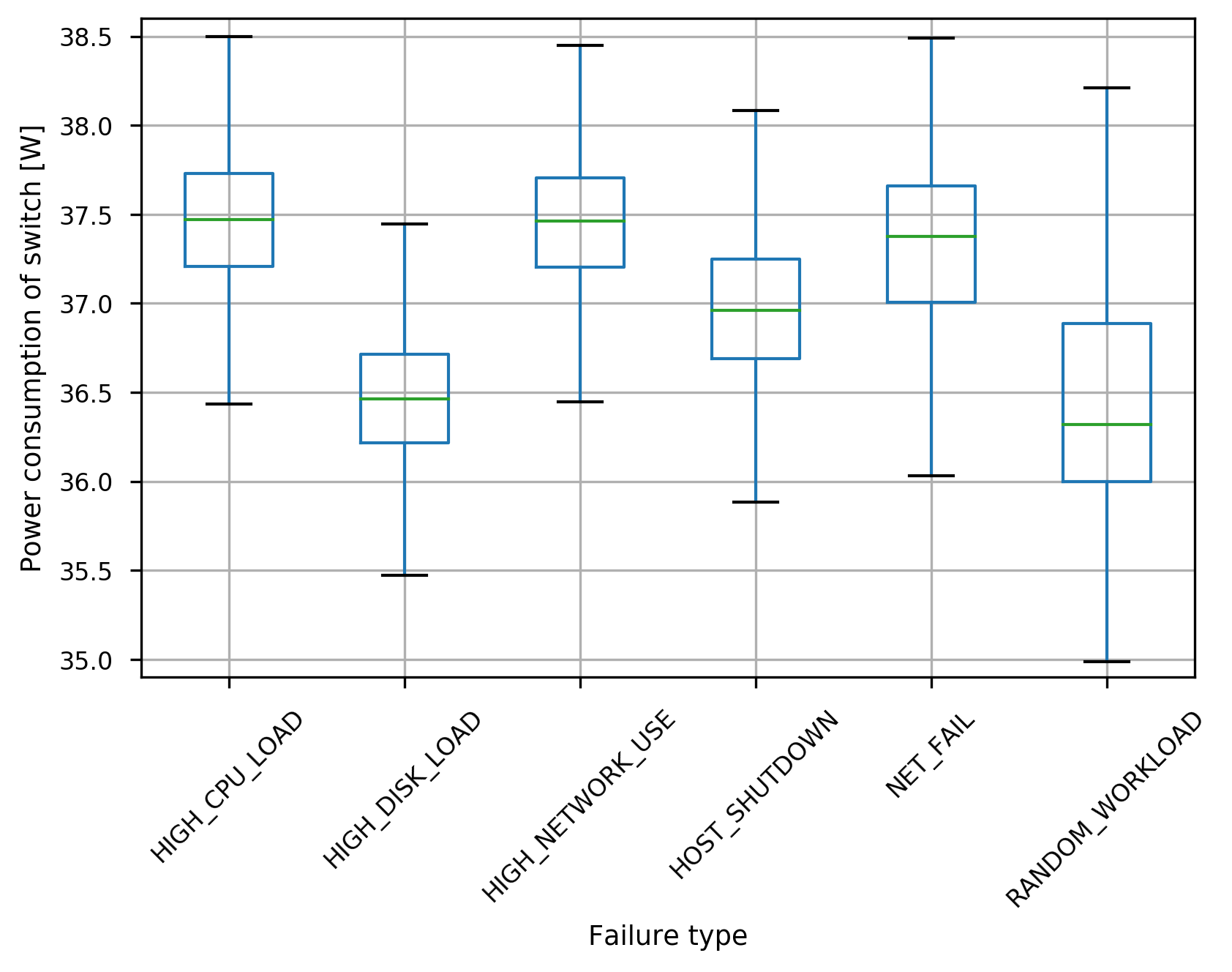}
 \centering
 \caption{An impact of different failure types on the power consumption of a switch. In random workload no failures are injected.}
 \label{fig:switch_power}
\end{figure}

The symptoms of failures have an understandable impact on system metric values. As we mentioned before, we include power metrics of the servers and the switch. Regarding the switch power, we can observe different peaks and power values depending not only on the network transfer but also on the connection and disconnections. In Figure~\ref{fig:switch_power} we present the switch power distribution depending on the injected failure, and the referential distribution for the system running random workload without any failure injected. We can observe that different power consumption values characterize different failures. These distributions increase the quality of failure classification in similarity evaluation. For instance, high disk load manifests in a low switch power consumption, while high network use manifests in significantly higher median value.

To evaluate the quality of the root cause classification, we use \textit{f1-score} metric that is defined as follows.
\begin{definition}
f1-score is the harmonic mean of precision and recall: $f_1 = \frac{2}{\frac{1}{recall} +\frac{1}{precision}} = 2 \cdot \frac{precision \cdot recall}{precision + recall}$
\end{definition}

\subsection{Evaluation: Leveraging logs for root cause classification}

 \begin{figure}[ht]
 \includegraphics[width=0.4\textwidth]{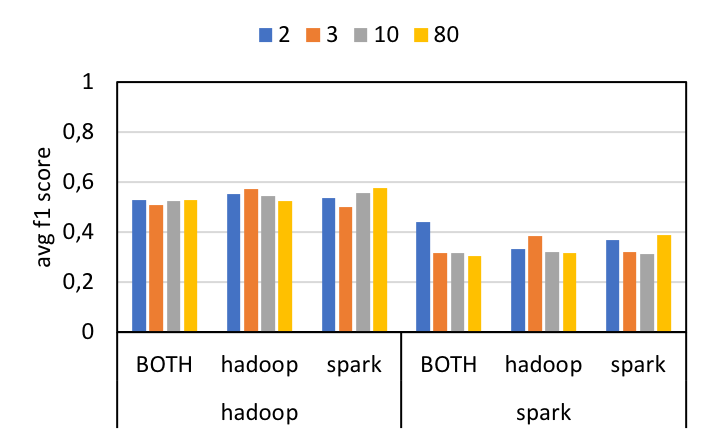}
 \centering
 \caption{Plot presenting the quality of root cause classification depending on the number of dimensions used in Word2Vec model, and the training vocabulary source. Log window length: 30s.}
 \label{fig:word2vec_eval_vec_size}
\end{figure}

\begin{figure*}[ht]
 \includegraphics[width=0.9\textwidth]{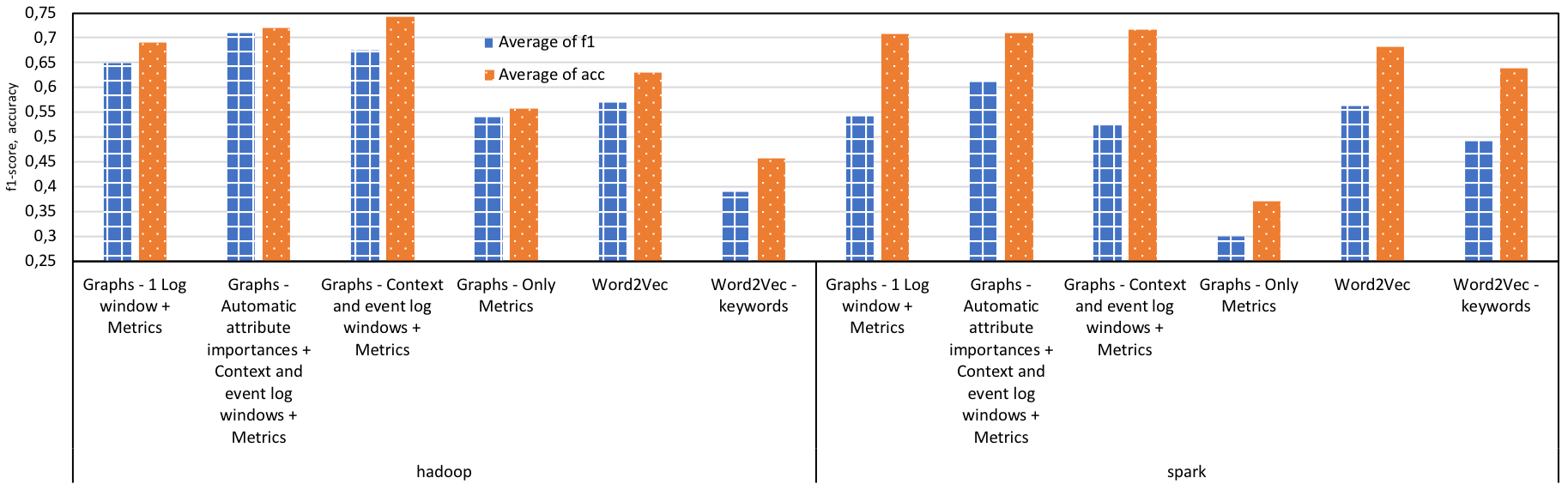}
 \centering
 \caption{Plot presenting root cause classification quality depending on the mechanism used. Average \textit{f1-score} is calculated from all of the injected failures. The proposed framework performs better than state of the art solutions (Word2Vec).}
 \label{fig:plot_results_all}
\end{figure*}

Firstly, we evaluate different methodologies and their configurations for the use of logs for the classification task. In the evaluation, we present the result of solving the following problems.
\begin{itemize}
 \item \textbf{Model training vocabulary.} We fit Word2Vec models using different vocabulary. It can be a specific vocabulary for a particular domain or a general dictionary \eg English one. For instance, we can train such a model with logs from Spark cluster, and use this model to vectorize Hadoop logs.
 \item \textbf{Model size.} We evaluate different numbers of dimensions of a vocabulary space (vector size). 
 \item \textbf{Key terms extraction.} We compare the performance of use of the whole available log entries with the key terms describing the system state. 
 \item \textbf{Log window length.} Size of the window is a trade-off between generalization of logs and capturing precise event information. Taking to much text can fuzzify the meaning of the event, and opposite, taking too little text can mangle an analyzed system state. We evaluate different window lengths for both event and context windows.

\end{itemize}

Firstly, we test how different sources of vocabulary and model size impacts the quality of the classification task. We create Word2Vec models with the process described in Subsection~\ref{subsection:logs} evaluating different vector size and vocabulary used for model training. In Figure~\ref{fig:word2vec_eval_vec_size}, we see average \textit{f1-scores} of the failure classification for the two use cases: the Spark and Hadoop cluster. We present only the best results achieved during the evaluation of different log window sizes. Also, we present summarized results of vector size evaluation. For vector sizes between 3 and 80 \textit{f1-scores} does not change much. In the Figure~\ref{fig:word2vec_eval_vec_size}, the inner groups stand for the source of the vocabulary used for model training. As well as for Hadoop and Spark, the classification performs the best when the same vocabulary is used for model training and vectorization. For both use cases, the models perform well with small vector sizes - 3 for Hadoop and 2 for Spark.

In the next step, we test different approaches of extracting information from logs and representing it in graphs. For the first approach, we use Word2Vec, as described above. In the second approach, we use SGRank~\cite{danesh2015sgrank} algorithm to extract key terms which best describe a system state. This algorithm combines statistical methods, \eg TF-IDF, with graph-based approaches of key terms. In Figure~\ref{fig:plot_results_all}, we confirm that using the whole text is the best method to represent the log meaning~\cite{word2vec_Mikolov:2013}.

\subsection{Evaluation: Root cause classification via similarity of weighted graphs}
In this subsection, we present the results of evaluation of the root cause classification. We test four different configurations of the proposed framework and compare them with the state of the art methods. We show how augmenting the dataset used for the classification task improves its performance. In Figure~\ref{fig:plot_results_all}, we present the results of the evaluation: average \textit{f1-score} and accuracy. Average \textit{f1-score} is calculated over all of the injected failures. In evaluations where it is emphasized that we use automatic attribute importance assignment, we utilize both similarity function based on distribution and automatic weight calculation. In others, we use equal weights in a graph.

We can see that the proposed framework that contains context and event log window and automatic attribute importance calculation performs better than state of the art methods. Considering performance for two use cases, graphs with automatic weights reveals the best performance. Regarding the Hadoop use case, accuracy reaches $0.72$, and \textit{f1-score} reaches $0.71$. As for the Spark use case, \textit{f1-score} is a little bit lower $0.61$ and accuracy $0.71$. Note that in the case of Hadoop, adding the automatic weight calculations lowers the \textit{f1-score}. Most probably it is because of that, the resource usage does not need to follow normal distribution~\cite{hadoop_workload_2013}, which we use as an estimator in the evaluation.

\begin{figure}[ht]
 \vspace{-10pt}
 \includegraphics[width=0.4\textwidth]{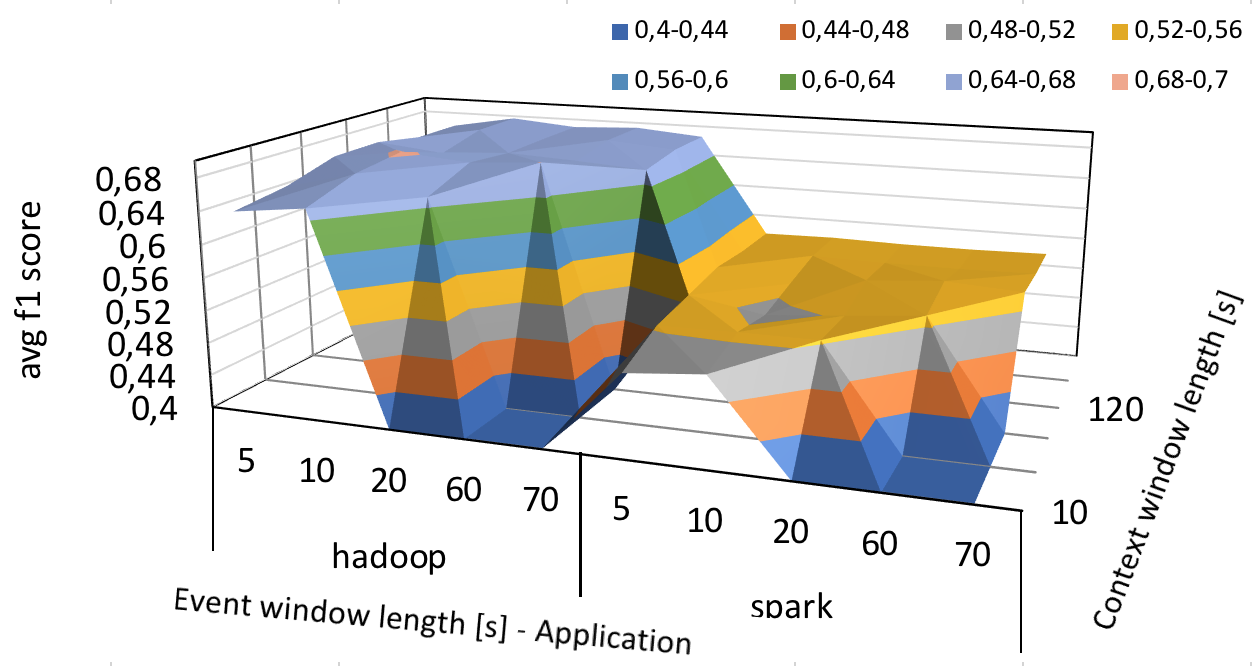}
 \centering
 \caption{Plot presenting quality of failure classification via graphs with equal weights depending on the log window sizes. Average \textit{f1-score} is calculated from all of the injected failures.}
 \label{fig:window_sizes_eval}
\end{figure}

We evaluate the proposed framework for different event and context window lengths. In Figure~\ref{fig:window_sizes_eval}, we present detailed results of this evaluation. The performance changes smoothly, there are local maxima of \textit{f1-score}. These maxima show balance points between log generalization and extraction of precise information about a particular event. The greater is log window length, the more fuzzified information about an event is held in analyzed window.

\begin{figure}[ht]
\vspace{-10pt}
 \centering
 \subfloat[Word2Vec model with parameters reaching the maximum quality, chosen from Figure~\ref{fig:word2vec_eval_vec_size}. Log window length:30s ]{{\includegraphics[width=0.48\textwidth]{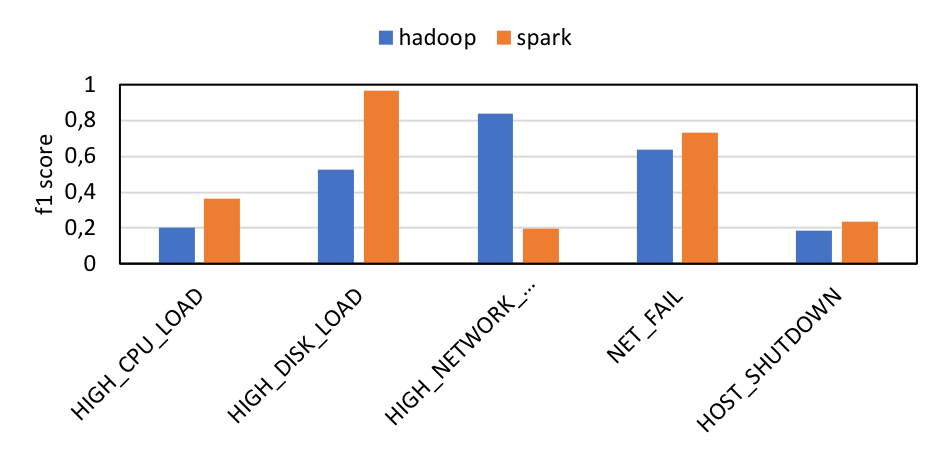} }}\hfill\subfloat[Automatically weighted graphs. Context window length: 30s, event window length: 10s, metrics window length: 120s]{{\includegraphics[width=0.48\textwidth]{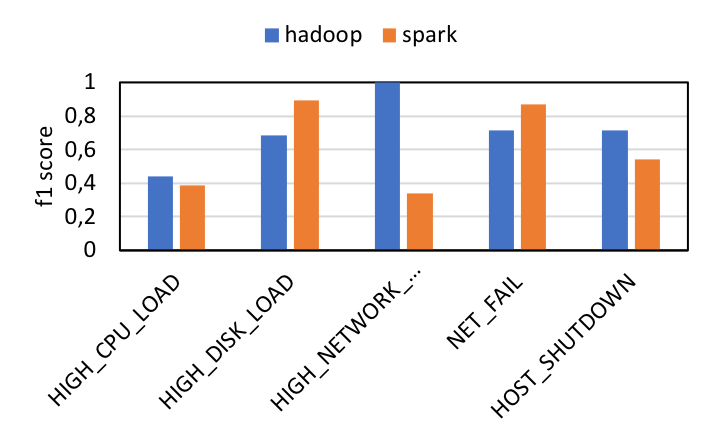} }}
 \caption{Plots presenting quality of failure classification for Word2Vec and the proposed framework.}
\label{fig:eval_window_size}
\end{figure}

In Figure~\ref{fig:eval_window_size} we present detailed evaluation results for each of the injected failures. We compare the use of logs with the proposed framework comprising automatically weighted graphs. The proposed framework performs significantly better than Word2Vec, especially with the classification of \textit{high CPU load} and \textit{host shutdown}. There is no observable difference in the performance of the proposed framework when used for the Spark or Hadoop use case. The exception is \textit{high network transfer}, which is classified well only for Hadoop by both Word2Vec and the proposed framework. \textit{High network transfer} manifests in characteristic log entries for Hadoop, and for Spark only in network metrics. Also, it is important to emphasize that, received results come from similarity evaluation of graphs created automatically without any weight adjustment by a human.

\section{Evaluation and experiments: Cross-system diagnostics - transferring knowledge}\label{section_evaluation_cross_domains}

\subsection{Experimental environment}
\begin{figure}[ht]
 \includegraphics[width=0.48\textwidth]{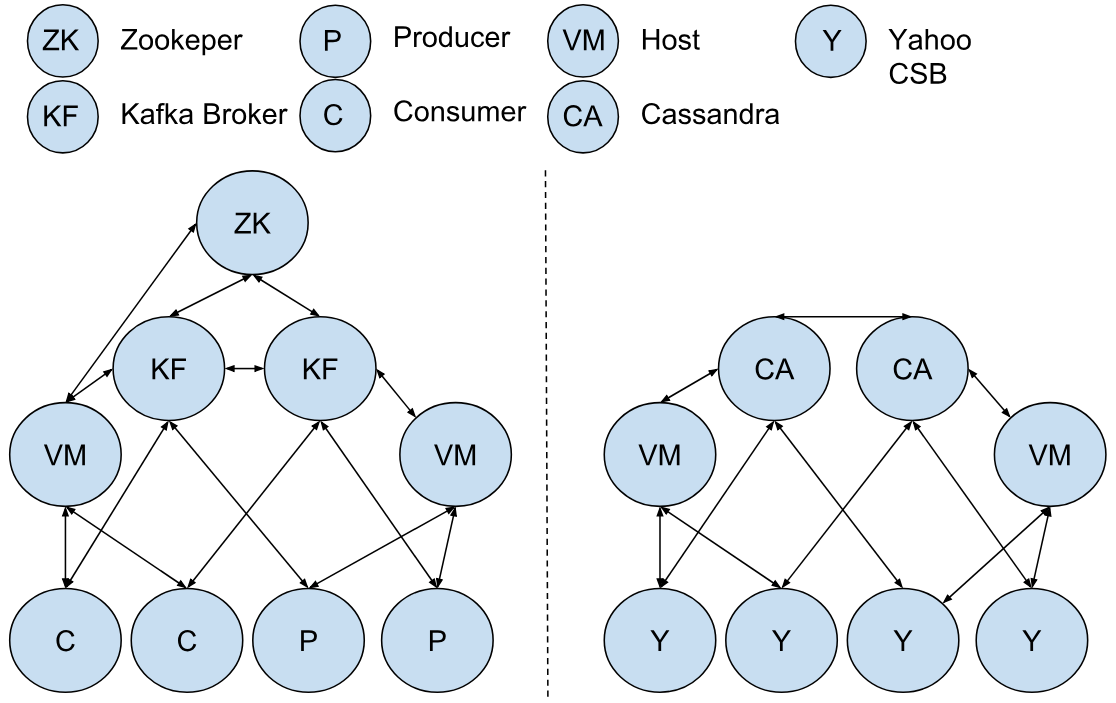}
 \centering
 \caption{A simplified version of the graph representations we use for the microservice architectures. On the left the Kafka architecture with a Zookeeper instance coordinating the brokers and producers and consumers using the message queue. On the right a Cassandra cluster with the YCSB clients. Notice how the VMs are connected to the containers they are hosting through edges that represent this relationship.}
 \label{fig:microservices}
\end{figure}

In this section, we evaluate our approach in a more cloud-oriented environment, by running microservice architectures made up of containers. We use Grid'5000 a customizable testbed that provides access to different computing resources and infrastructures. We deploy a cluster of 7 virtual machines with 16GB of RAM and four cores. We install DC/OS\footnote{https://dcos.io/} on these machines, a container orchestration tool that will allow us to deploy the microservice architectures. The setup is 1 master node, 1 public node, and 5 private nodes. Additional information about DC/OS parts can be found in their website\footnote{https://mesosphere.com/}. We use two additional representative Big Data architectures to perform root cause analysis with them. The first one is a Cassandra deployment with 5 Cassandra containers that are going to be continuously queried by 10 containers with Yahoo Cloud Service Benchmark~\cite{YCSB} installed. The second one is a Kafka architecture, in which we have 5 brokers, 10 producers that push messages to the Kafka cluster and 10 consumers that read those messages. Additionally, the Kafka brokers need a Zookeeper~\cite{zookeeper} instance to coordinate them. A simplified version of the graph representations we use for these deployments is shown in Figure~\ref{fig:microservices}. Note that these two architectures are very similar with a decentralized cluster of servers or brokers that interact with each other and clients that read or write data into this cluster. This scenario is a suitable one for our knowledge transfer approach since failures that happen in one system will have a similar effect if they also occur in the other one. 

\subsection{Methodology}
Regarding the failures, we injected them in both the hosts and the containers. For the hosts, we use the same high CPU, high disk, and high network transfer anomalies as in the Spark scenario to stress the machines. For the containers, we pause them through \texttt{docker pause} instead of using host shutdown and network failure. We do so because a container cannot be physically disconnected from the network as a host would. The anomalies are injected six times each, in one random element of the architecture for 120 seconds.

\subsection{Evaluation: Cross-system diagnostics}
\begin{figure}[ht]
 \includegraphics[width=0.4\textwidth]{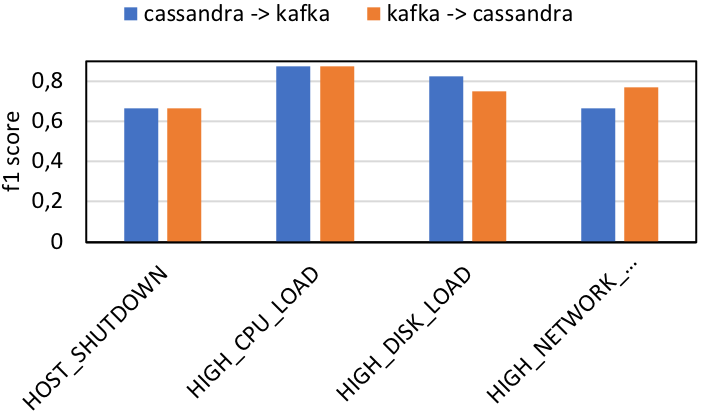}
 \centering
 \caption{Plot presenting results of cross-system diagnostics via finding the nearest graph representing an anomalous state of a system. Results of two cases are presented. 1) \textit{Source system}: Cassandra, \textit{target system}: Kafka; 2) \textit{Source system}: Kafka, \textit{target system}: Cassandra. Average \textit{f1-score} and accuracy: 1) 0.76, 0.77; 2) 0.77, 0.77.}

 \label{fig:cross_diagnostics_results}
\end{figure}

We present detailed results of the evaluation in Figure~\ref{fig:cross_diagnostics_results}. Average \textit{f1-score} is $0.77$ in case of using Cassandra as a source system and Kafka as a target one. In the reversed configuration, the result is $0.76$. Note that the scores of the cross-system diagnostics are better than the first evaluation of the framework, due to the different number of types of the injected failures. Both quality results are approximately equal, thanks to the symmetry of similarity function. The small difference is caused by the task of finding the nearest graph (a one with the highest similarity number). This operation is not always symmetric. Considering that two systems are different, in their topology, behavior, and logs, the results are showing high performance of the proposed framework.

\section{Discussion and Conclusion}\label{section_discussion}
In this paper, we proposed a framework for finding the nearest failure cause via similarity evaluation of weighted graphs. The framework is aimed to diagnose one system when the knowledge about failures is acquired from another system with a different structure. An example would be a new system that has just started operating, it fails, and it is hard to diagnose it. Also, the proposed framework aims to facilitate knowledge transfer between systems and operators. Firstly, we described the whole framework and its contributions. The most significant contributions are automatic calculations of metric weights, integration of logs with system topology and metrics into graph representation of a system, and leveraging historical metric values for similarity calculations. Then, we evaluated the proposed framework in total with four different systems. We inject common anomalies and failures, such as hardware overload, node crash, and network disconnections. In the first evaluation section, we use Spark and Hadoop clusters. We confirm the quality of root cause classification that achieves average \textit{f1-score} of 0.71 for Hadoop and 0.61 for Spark. These results show that the framework outperforms state of the art methods. In the second evaluation, we utilize a cloud environment of containers. We evaluate cross-system diagnostics via knowledge transfer. That means diagnosing a target system when knowledge about failure causes and anomalous states is known only from a source system. We run two scenarios: Kafka acting as the source system and Cassandra as the target one and vice versa. Cross-system diagnostics reaches average \textit{f1-score} of $0.77$. The achieved results confirm that the proposed framework, and in particular its ability of knowledge transfer, allows reaching the state of self-manageable IT systems.

In the next stage of research we can focus on:
\begin{itemize}

\item Evaluation of the framework on the real large-scale environments. Also, an integration of the proposed framework with the proactive failure prevention system might be useful.

\item There might be interesting research performed on knowledge transfer framework integrated with knowledge exploration solutions. Such a system could automatically mine knowledge on failures from parts of the system.

\item Another vital issue to consider in the future work is an automatic taxonomy construction. Then the knowledge transfer would be much more automated. 

\item Aspect of explainable knowledge transfer in cross-system diagnostics.
\item Distinguishing random errors, and the ones which are critical for the future system performance and reliability. 
\item Mechanism for automatic propagation of weights for anomalous regions inside graphs
\item Research in the field of predicting failures with use of transfer learning.

\end{itemize}

\section*{Acknowledgment}
This research is supported by the BigStorage project (ref. 642963) funded by Marie Sk\l{}odowska-Curie ITN for Early Stage Researchers, and it is a part of a doctorate at UPC. Special thanks to German Climate Supercomputing Center (DKRZ, Hamburg, Germany) and Grid'5000 (Inria, France) for the access to computing resources for the experiments.

\bibliographystyle{IEEEtran}

\end{document}